# The Brachistochrone: An excellent problem for all levels of physics students.

John A. Milsom

Department of Physics, University of Arizona, Tucson, AZ.

The classic brachistrochrone[1] problem is standard material in intermediate mechanics. Many variations exist[2,3,4] including some accessible to introductory students.[5,6] While a quantitative solution isn't feasible in introductory classes, qualitative discussions can be very beneficial since kinematics, Newton's Laws, energy conservation and motion along curved trajectories all play a role. In this work, we describe an activity focusing on a *qualitative* understanding of the brachistochrone and examine the performance of freshmen, juniors and graduate students. The activity can be downloaded at

https://w3.physics.arizona.edu/undergrad/teaching-resources.

Simply stated, the brachistochrone is the shape of the frictionless path that minimizes the travel time between two points in a constant gravitational field. The solution is a cycloid (see Fig. 1). We focus here on the case where the point mass is released from rest. Under these conditions, the solution is also a tautochrone so it has the remarkable property that the descent time to the bottom is independent of the starting location.

Four student groups worked this activity. Groups I (350 science/engineering majors) & II (62 physics/astronomy/honors students) were taking introductory calculus-based mechanics. They worked in pairs and had already covered kinematics, Newton's Laws, circular motion and were partway through work/energy. Group III was thirty 5th semester physics/astronomy majors taking Lagrangian mechanics. They worked on it individually before seeing the quantitative solution.

Lastly, Group IV was seven graduate students who facilitated the activity for Groups I & II after answering the questions themselves. They monitored the students' work, answered questions and asked probing questions when they saw mistakes. Since they were each running sections with twelve student pairs, they had minimal time to spend with each pair.

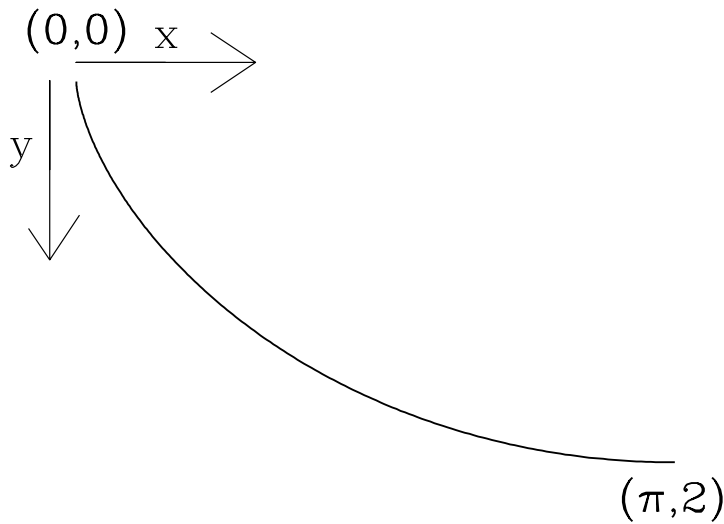

Figure 1: The cycloid solution to the brachistochrone. The radius of the rolling ball which "traces" this cycloid is $R = 1$ m.  Masses released from rest at any point along the path reach the bottom at the same time. Note that $y$ increases downwards.

## The questions and the students' performance:

In this section, we state the questions and discuss each group's performance. Please refer to the detailed performance data in Table I while reading this. Note that the online activity has a 5th question not discussed here since our students didn't get to it given time constraints.

Before the students began, we used a handmade brachistochrone to demonstrate two things. First, since the model also has a straight path, we demonstrated how much faster the cycloid path is. Second, since the model has a descending and ascending path we released masses on opposite sides at different heights and they collided at the very bottom. This demonstrated the tautochrone property.

**Question 1**: *"Explain how it is possible for the mass to start anywhere on the path and still reach the very bottom in the exact same amount of time."*

Here are student answers from Groups III, I, II and IV respectively:

- *The higher it starts, the greater its acceleration but it does have more distance to cover. If it starts lower, it has less acceleration but less distance to cover.*
- *An object that starts lower on the path begins with less potential energy which translates to a lower kinetic energy and velocity. Since it moves slower, it reaches the end at the same time as a faster object which started higher up.*
- *If it starts at a lower point, the ball will have less time to accelerate. Therefore, it does not matter where it starts, it will reach the bottom at the same time.*
- *Starting at a point closer to the bottom means there is less distance to travel but the average velocity will be lower.*

The first answer above is clearly a good one. While that student doesn't say "tangential acceleration" or "acceleration along the path", that student clearly gets it. While the second answer was relatively common (see Table I), it is inaccurate since one could make that comment about a straight track. Overall, the incorrect answers indicated numerous misconceptions. There were inconsistent answers (answer three), students who think the acceleration is constant, mixing up velocity and acceleration, thinking that centripetal acceleration matters, etc. The fourth answer above lacks a complete explanation. Many students gave similar incomplete answers. When prodded, some did understand the solution but didn't realize their answer was insufficient.

| Percentage who: | Group I % | Group II % | Group III % | Group IV % |
|---|---|---|---|---|
| Answered question 1 correctly / Used energy instead. | 22 / 14 | 26 / 32 | 20 / 27 | 57 / 0 |
| Scored 3 / 2 / 1 on $x(t)$ graph. | 51 / 20 / 29 | 75 / 16 / 9 | 71 / 25 / 4 | 57 / 43 / 0 |
| Scored 3 / 2 / 1 on $v_x(t)$ graph. | 9 / 36 / 55 | 34 / 16 / 50 | 32 / 50 / 18 | 71 / 29 / 0 |
| Scored 3 / 2 / 1 on $a_x(t)$ graph. | 17 / 11 / 72 | 47 / 9 / 44 | 39 / 15 / 46 | 86 / 0 / 14 |
| Scored 3 / 2 / 1 on $y(t)$ graph. | 29 / 23 / 48 | 13 / 54 / 33 | 20 / 48 / 32 | 71 / 14 / 14 |
| Scored 3 / 2 / 1 on $v_y(t)$ graph. | 17 / 27 / 56 | 42 / 21 / 38 | 25 / 42 / 33 | 43 / 57 / 0 |
| Scored 3 / 2 / 1 on $a_y(t)$ graph. | 13 / 26 / 62 | 21 / 33 / 46 | 4 / 42 / 54 | 0 / 86 / 14 |
| Thought $a_x$ / $a_y$ was constant. | 37 / 43 | 19 / 33 | 31 / 42 | 0 / 0 |
| Thought $a_x = 0$ | 8 | 3 | 15 | 0 |
| Drew graphs with consistent derivatives. | 61 | 63 | 62 | 82 |
| Drew $n$ vertical | 10 | 0 | 0 | 0 |
| Accounted for force orientation and / or acceleration when drawing $n$. | 7 / 39 | 13 / 52 | 4 / 50 | 43 / 43 |
| Had the relative magnitudes of $n$ and $mg$ consistent with $a_y$ at the bottom of the cycloid | 27 | 38 | 38 | 71 |

Table I: Summary of each group's performance. Students in Groups I and II worked in pairs while those in Groups III and IV worked individually.

**Question 2**: "Graph $x(t)$, $v_x(t)$, $a_x(t)$ qualitatively on the axes provided here and explain why each graph has the shape you have drawn. (Assume the mass was released from rest at the top.)"

**Question 3:** Same as **2** but for $y(t)$, $v_y(t)$, and $a_y(t)$.

The difficulties students have in interpreting and drawing graphs are well-known.[7,8] These are particularly challenging since the radius of curvature continuously changes. We *coarsely* graded all six graphs on a scale of 1-3. Roughly, 1 means mostly incorrect, 2 means partly correct/incorrect and 3 means mostly correct. We also separately checked whether $v_i = \frac{dr_i}{dt}$ and $a_i = \frac{dv_i}{dt}$.

Panels a-c in Fig. 2 have the correct graphs which instructors can provide to students later. They exhibit features that may not be widely known. Introductory textbooks generally illustrate the connection between uniform circular motion and simple harmonic motion.[9] Since this brachistochrone solution is identical to following a point on the edge of a sphere rolling without slipping, it is uniform circular motion coupled with a translation. Thus, the vertical motion is still simple harmonic motion and $a_y$ varies from $g\hat{y}$ to $-g\hat{y}$.[10] Additionally, the accelerations in panel c are just sines and cosines with amplitude $g$ so $|\vec{a}| = g$ at all times. Therefore, the acceleration has a constant magnitude but a continuously changing direction (just like uniform circular motion).[11]

One could devote an entire paper to these results, but here we highlight specific issues.

- Only the $x(t)$ graph was drawn well by all groups.
- Many students think $a_x$ and/or $a_y$ is constant. For example, one Group III student wrote "*The acceleration is g since gravity is the only force in the y-direction.*"
- Some students think $a_x = 0$.
- Most graphs were consistent with $v_i = \frac{dr_i}{dt}$ and $a_i = \frac{dv_i}{dt}$. While some students checked this, it was rare without us prompting it. These were frequently correct "by accident" since they had constant accelerations and knew the corresponding graphs.
- Only one Group III student realized that $a_y$ had to flip sign.

- Since this activity was used later in the semester than typical kinematic graphing exercises, students employed multiple graphing techniques. In addition to considering the endpoints and the intervening shape, some used free body diagrams to determine the acceleration. In a few cases, using multiple techniques helped them find mistakes.
- All seven Group IV graduate students had $a_y = 0$ at the very bottom. While six of them realized that $a_y$ flipped sign during the descent they had it end at zero.[12]

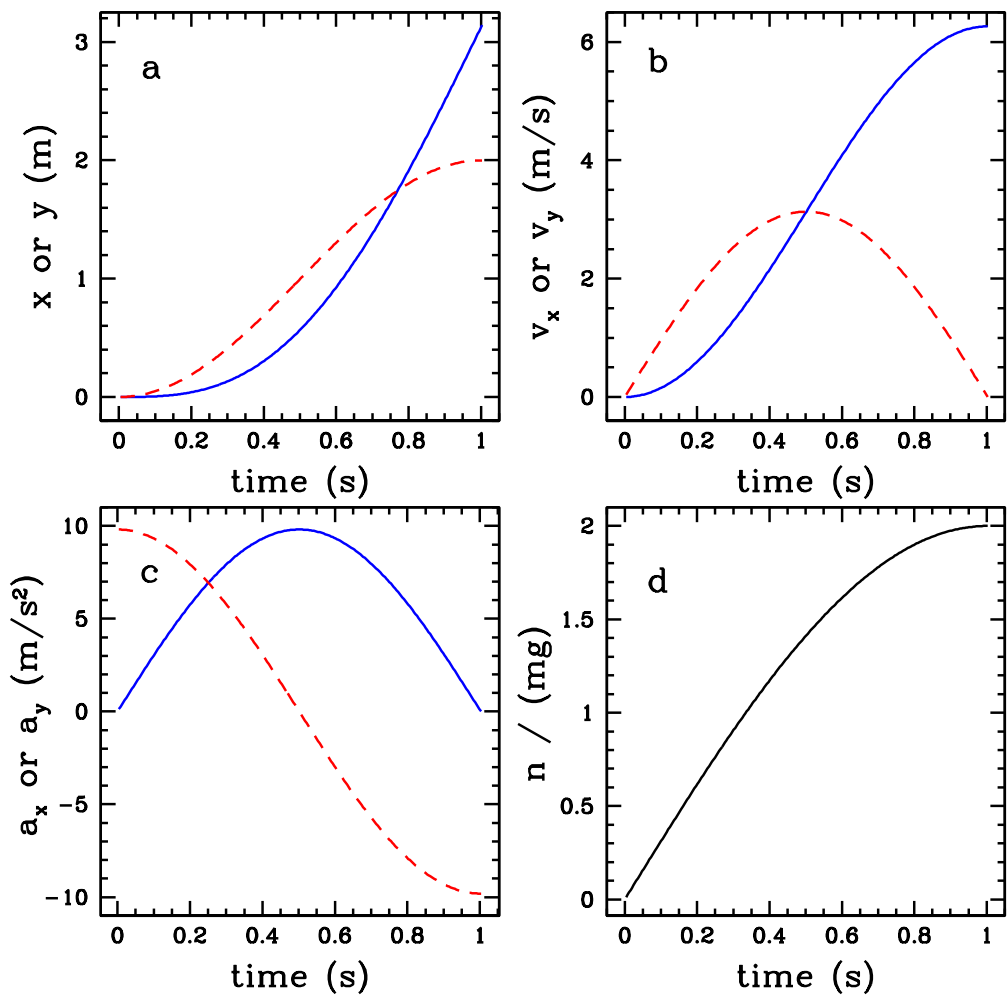

Figure 2: The panels (a-d) illustrate $x(t)$ (solid/blue) and $y(t)$ (dashed/red), $v_x(t)$ (solid/blue) and $v_y(t)$ (dashed/red), $a_x(t)$ (solid/blue) and $a_y(t)$ (dashed/red), and $n(t)/(mg)$ respectively when the mass is released from rest at the top. The total descent time is $\pi\sqrt{R/g} \simeq 1.00$ s.

**Question 4**: *Let's now consider the normal force exerted by the track on the mass. Discuss how you expect its magnitude to change as the mass slides down the track. Explain your reasoning. Draw qualitatively correct free body diagrams at each of the six points in the figure. Hint: three different factors influence its magnitude.*[13]

How do we expect $n$ to change? This is similar to analyzing the normal force for a mass sliding down a hemispherical bowl (or to analyzing the tension for a simple pendulum). The normal force increases during the descent because the angle between $\vec{n}$ and $m\vec{g}$ increases from $90°$ to $180°$ so $\vec{n}$ must overcome an increasing component of gravity perpendicular to the path. Additionally, the speed is increasing so the required centripetal acceleration also increases. However, for the cycloid there is a continuously increasing radius of curvature. This could either decrease the rate at which $n$ increases or possibly make $n$ decrease. We graph $n(t)/(mg)$ in panel d of Fig. 2 and see that overall $n$ does continuously increase.

Figure 3 illustrates *quantitatively* correct free body diagrams which could be provided to the students afterwards. Readers may be surprised how quickly $n$ becomes larger than $mg$ due to the rapidly increasing speed along the initially steep trajectory and the small radius of curvature there. By the second point, $|\vec{n}| \simeq |m\vec{g}|$. Since $\vec{a}_y = -g\hat{y}$ at the bottom, Newton's Laws dictate that

$n = 2mg$. For larger cycloids, $n$ remains $2mg$ since the radius of curvature and the mass's kinetic energy both increase linearly with the cycloid height.[14]

In checking their work, we primarily examined their thought process. Summarizing their performance:

- Very few students accounted for the changing force orientation *and* the changing centripetal acceleration.
- Only one Group III student did this correctly. This illustrates the importance of students seeing content multiple times. Since their Lagrangian mechanics course hadn't (yet) covered anything reminiscent of centripetal acceleration, this issue eluded them. Conversely, some "freshmen" in Groups I & II recognized and accounted for the similarity to circular motion.
- A few students argued that $n$ should increase and then decrease due to the flattening trajectory.

- Some Group I students drew $n$ vertical everywhere.
- For most students, the relative magnitudes of $n$ and $mg$ at the very bottom were inconsistent with their corresponding value for $a_y$. While not surprising, this highlights the difficulty in getting students to consider the connections between questions.

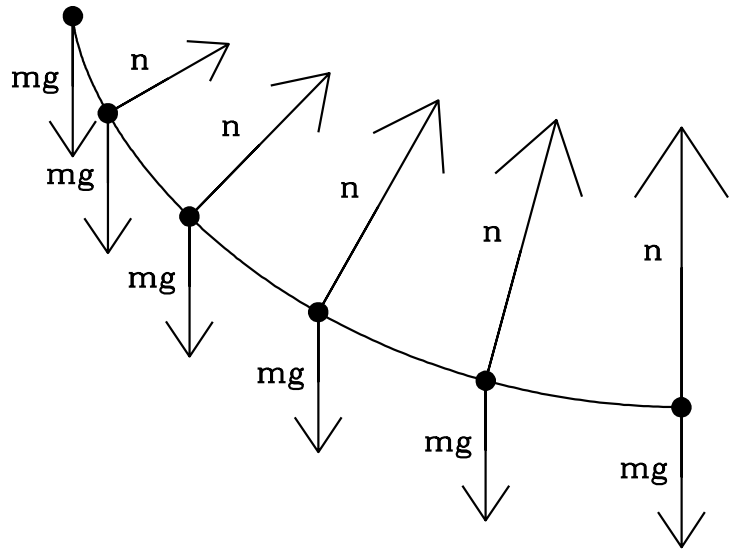

Figure 3: *Quantitatively* correct free body diagrams at specific points. The point mass is superimposed on the trajectory. The cycloid slope is infinite at the top so the normal force is zero there.

## Summary

This activity requires a qualitative analysis of the famous brachistochrone problem and helps students review approximately half of a typical introductory mechanics course. It will pique their interest as they tire of never-ending incline planes and Atwood machines. While clearly challenging, it leads to beneficial student discussions and requires the content of multiple textbook chapters.

As expected, Table 1 indicates that the freshmen science/engineering majors in Group I had the worst performance while the graduate students in Group IV had the best. However, the freshmen physics/astronomy/honors students in Group II very slightly outperformed the junior physics/astronomy majors in Group III.[15] It is obvious that advanced students would benefit from this refresher.

Given their disappointing performance, we have several recommendations/suggestions.

- We *strongly* recommend the introductory students work in groups of four and that there are sufficient learning assistants to help them.
- Fifty minutes is needed to work through this.
- Most students clearly need significant guidance. While we heard many interesting discussions, they frequently reached the wrong conclusion(s). The activity's last page is an instructor supplement including hints which could/should be given to improve performance and learning. For example: Can your understanding of circular motion help you draw these free body diagrams? Instructors can decide what hints are appropriate for their students and also provide their own. They can also decide whether to provide them up front to the entire class or to individual groups as needed.
- Faculty may choose to have students draw the free body diagrams before drawing the kinematics graphs. This may alleviate the large number of incorrect acceleration graphs.
- We recommend providing the correct graphs at the end since this led to further discussions.

The author thanks professors Alex Burant, Brian LeRoy, John Schaibley, and Charles Wolgemuth for having their students work this activity. The author also thanks the anonymous referees for their comments and suggestions.